# Effective Metaheuristic Based Classifiers for Multiclass Intrusion Detection

Zareen FATIMA[1], Arshad Ali[2]

[1]Department of Computer Science, University of Lahore, Lahore, Pakistan,

ORCID iD: https://orcid.org/0000-0002-1198-5883

[2]FAST School of Computing, NUCES, Lahore Block B Faisal Town, Lahore, 54770, Pakistan,

ORCID iD: https://orcid.org/0000-0003-0562-7403

**Abstract:** Network security has become the biggest concern in the area of cyber security because of the exponential growth in computer networks and applications. Intrusion detection plays an important role in the security of information systems or networks devices. The purpose of an intrusion detection system (IDS) is to detect malicious activities and then generate an alarm against these activities. Having a large amount of data is one of the key problems in detecting attacks. Most of the intrusion detection systems use all features of datasets to evaluate the models and result in is, low detection rate, high computational time and uses of many computer resources. For fast attacks detection IDS needs a lightweight data. A feature selection method plays a key role to select best features to achieve maximum accuracy. This research work conduct experiments by considering on two updated attacks datasets, UNSW-NB15 and CICDDoS2019. This work suggests a wrapper based Genetic Algorithm (GA) features selection method with ensemble classifiers. GA select the best feature subsets and achieve high accuracy, detection rate (DR) and low false alarm rate (FAR) compared to existing approaches. This research focuses on multi-class classification. Implements two ensemble methods: stacking and bagging to detect different types of attacks. The results show that GA improve the accuracy significantly with stacking ensemble classifier.

## 1. Introduction

Due to rapid growth in computer networks and applications, networks security has become a major challenge in cyber security field. Intrusion detection plays a key role to detect different attacks in computer networks [1]. The goal of an Intrusion Detection System (IDS) is to monitor network traffic to identify any malicious activity that can breach confidentiality and integrity information. Accordingly, the IDS alerts the network administrator or system about such activities or attacks [2]. An attack is a malicious activity or unauthorized access that affects the network system and can get access to confidential data. Typically, there are five major categories of attacks: Denial of service (DoS), brute force attack, Probe attacks, user to root, remote to local attack. An IDS is capable of responding to different attacks and reporting them accordingly. Typically, an IDS is divided into two categories i.e., network-based intrusion detection system (NIDS) and host-based intrusion detection system (HIDS). NIDS monitors inbound and outbound network traffic. NIDS captures whole network traffic and evaluate the separate packets to detect abnormal activities [3]. HIDS monitors computer or host system to detect malicious activities. HIDS detect intrusion by monitor the system call, application activities, scheduling process, login attempts and system configuration on a specific machine. If any malicious activity or changes occur in the system, it generates an alarm [2,3,4]. Moreover, there are three general intrusion detection types: anomaly-based, misuse (signature-based) and hybrid-based. Anomaly based Intrusion detection approaches relay on the attacker behavior by determining the user profile and uses as baseline to define normal user activities. When compare attacker activities with baseline and it generates an alarm to aware the network administrator. Anomaly detection can detect unknown attacks but there with high false positive rate [3,4]. The misuse intrusion detection method stores the attack pattern or signature and compares this pattern to network traffic. If it's matched, it generates an alarm of intrusion. This method can detect only known attacks [3,4]. Hybrid method is a combination of both anomaly and signature-based intrusion detection methods and can detect both known and unknown attacks.

*Corresponding author: arshad.ali1@nu.edu.pk

One of the major challenges to detecting attacks or malicious activities is analyzing large amounts of data. Intrusion detection systems face big data challenges [2]. Therefore, feature selection plays a key role in reducing big data problems by selecting the most relevant features [2]. Features selection is a machine learning technique that selects the most relevant subset of the original features set that achieves high detection accuracy as compared to the original features set [5]. By reducing the dimensional of datasets to remove the irrelevant and redundant features, machine learning algorithms can make more efficient classification predictions. The selection of best feature subsets can reduce training and testing time, improve detection rates, reduce false alarms rate, and create lightweight datasets that can be used to build IDS for real-time and online attack detection [2]. There are three comprehensive methods for features selection such as filters, wrappers, and hybrid-based methods [6].

Filter based feature selection approach uses independent algorithms to select features and applies external algorithms to evaluate the performance of selected features [6,7]. This approach uses statistical measure to evaluate the relationship between each input variable and target variable. It selects input variables that have a strong correlation with the target variable and considers the most relevant features. This feature method can be easily applied as it does not use learning algorithms in the feature selection procedure [1]. Wrapper method "wrapped" around the learning algorithm. This selection method uses a learning algorithm to select the important features subsets. The Wrapper algorithm uses a search algorithm to evaluate the significance of different feature subsets, where feature subsets' worthiness is evaluated by a learner [6]. The Wrappers method is computationally expensive but, in comparison to other feature selection approaches, it is more precise [1,2]. Hybrid methods combined the benefits of filter-based methods and wrapper-based methods to obtain a best feature selection subset by using a learning algorithm.

In this research, we suggest a wrapper-based FS technique based on the Genetic Algorithm (GA) that generates optimal feature sets by using the Naïve bayes (NB) ML algorithm as its fitness function. To evaluate the performance of the suggested method, we use two intrusion detection datasets, UNSW-NB15 and CICDDoS2019. These datasets are latest and have most up-to-date attacks. The UNSW-NB15 dataset is widely used in feature selection methods evaluation, but CICDoS2019 is not commonly used.

The contribution and purpose of this research are as follows:

- We use recent and the most up-to-date attack datasets.
- Firstly, we select a GA-based feature selection method. To compute the fitness function, we used the Nave Bayes (NB) learning algorithm in the GA process.
- Secondly, for each selected feature set evaluation, we apply two ensemble-based classifiers, stacking and bagging.
- Finally, we compare our suggested method with existing methods. The results show a significant improvement in performance.

The reminder of the paper is organized as follows. Section 2 describes the related work. Section 3 demonstrates the suggested IDS methodology. Section 4 presents the experiments and discussion of the results. Section 5 provides the conclusion of this paper.

## 2. Related Work

This section gives an overview of relevant research works that used machine learning techniques in the domain of IDS. This section also provides an overview of several IDS frameworks and solutions.

In [8], the authors performed the features selection method on the UNSW-NB15 dataset. In this study, features were selected by using information gain obtained through XGBoost classifier. Features with a higher information gain score were considered more important features. They selected the 23 most important features and applied the XGBoost classifier to predict different attack types. On the testing dataset, they achieved 75.88% accuracy.

In [9], statistical and heuristic-based feature selection methods (forward selection and backward elimination) were used for features selection. These methods were performed on DARPA dataset. Resilient back propagation neural networks used for classification, it increased the accuracy and gave less training and testing time then tradition neural network. Heuristic and statistical based methods selected the features for each class type. The accuracy difference is very small or may not be statistically significant because the pattern of five class's size has huge difference.

In [7] Fisher Score algorithm was used to select the best features. They used CICIDS2017 datasets and classified the datasets as benign or DDoS by using SVM, KNN and Decision Tree (DT) algorithms. Fisher Score algorithm reduced features from 80 to 30. According to the order of importance, the "Fwd Packet Length Mean" feature considered the most important feature to detect intrusion. With 60% reduction in size of datasets, the success rate of KNN increased, DT accuracy did not change and SVM accuracy decreased.

In [10] author presented two stage classifier approach, in first stage incoming traffic divided in TCP, UDP or other protocols then it identifies the normal or attacks class. During preprocessing in first stage, features were selected by using information gain methods. In second stage, they used multiclass classification to identify the attack type. In the full dataset with all classes, it used Reduced Error Pruning Tree (REPTree) for classification. The experiments were performed on UNSW-NB15 and NSL KDD datasets.

In [11], researchers presented a filter-based approach by using XGBoost algorithm for features reduction. The experiments were performed on UNSW-NB15 dataset. XGBoost calculates the F-measure score for each feature of the given dataset, and high-scoring features are selected as important features. The number of features has been reduced from 42 to 19. The following ML algorithms were applied to the selected features subspace: ANN, LR, KNN, SVM, and DT.

The author of [12] created a new dataset, CICDoS2109, which contains the 11 DDoS attack types. This dataset has 11 classes in training dataset. It also applied the info gain method to select the important features. The authors applied different ML classifiers, ID3, RF, NB, and LR, to obtain accuracy. They evaluated the performance by using three metrics: precision, recall, and f1 scores.

In [13] authors applied ensemble feature selection methods on CICIDS2017 datasets and reduced features from 69 to 10. Gini importance, permutation importance and Drop-column importance are used as feature selection methods. Random forest classifier used for evolution. Permutation importance considered best methods between the accuracy of drop-column and the computational cost of Gini importance. The comparison of both original 69 features and 10 features determined the minor difference in F1-score 0.2 and false positive rate approximately 0.

Author of [14] proposed multi-objective feature selection method based on NSGA-II and logistic regression. Two schemes were used in this proposed work, for binary class dataset binomial logistic regression were used and for multi class dataset they used multinomial logistic regression. C4.5, RF, and NB classifier were used to evaluate the best subset features. For experiments CICIDS2017, NSL KDD and UNSW-NB15 datasets are used. Features Selected in NSL KDD datasets, 9-19 in binary class and 19 in multi class. In CICIDS2017 datasets selected features in binary class 7–25 and 33 in multi class. UNSW-NB15 datasets selected features 8-17 in binary and 11 in multi class. This paper gives better accuracy in binary class compared to multi class.

In [15], authors presented an IDS framework for features selection by using evolutionary genetic algorithm approach with SVM (GA-SVM). This paper proposed a new fitness function by using three evaluation parameters: FPR, TPR, and the number of selected features. SVM is also used for classifying the different types of attacks. KDD CUP 99 and

UNSW-NB15 datasets were used for experiments. The KDD CUP 99 dataset used five classes for the experiment: Normal, DoS, PROBE, R2L, and U2L. Furthermore, UNSW-NB15 dataset used seven classes: Normal, Fuzzers, Reconnaissance, Shell Code, DoS, Exploits, and Generic.

In [16], they proposed an IDS model for IIoT by using the Genetic Algorithm with random forest. The UNSW-NB15 dataset was used for the experiments. GA generated two feature vectors (FV), the first for binary classification and the second for multiclass classification. Ten feature vectors are generated for the binary class, and seven for multiclass. In the second step, logistic regression with tree-based algorithms (DT, RF, ET & XGB) is implemented on each FV. In binary classification, RF achieved a best test accuracy (TAC) and in multiclass Extra Tree (ET) achieved highest TAC.

In [17], the authors proposed a wrapper-based pigeon-inspired optimizer algorithm. This algorithm is based on cosine similarity. The proposed feature selection algorithm was tested on three datasets: KDDCUPP99, NSL-KDD, and UNSW-NB15. The proposed algorithm reduced the features of KDDCUP99 from 41 to 7, NSL-KDD from 41 to 5, and UNSW-NB15 from 49 to 5.

In [18] the authors proposed an IDS model based on feature selection methods and ensemble classifiers. First, they proposed a heuristic algorithm named CFS-BA. It selects the important feature subset based on the correlation between the input variable and the target variable. Then they proposed an ensemble classifier that combines C4.5, Random Forest (RF), and Forest by Penalizing Attributes algorithms. The evaluation was performed on three datasets: NSL-KDD, AWID, and CIC-IDS2017. The proposed algorithm reduced features of NSL-KDD from 41 to 10, AWID from 155 to 8 and CIC-IDS2017 from 80 to 13.

In [19], researcher suggested a hybrid based feature selection method with two ensemble classifier. Feature selection consists of particle swarm optimization (PSO), ant colony algorithm (ACO), and genetic algorithm (GA). For classification, they used two-level ensemble classifiers that were based on rotation forest (RF), (RF) and bagging (BG). RF and BG are used as a meta classifier, and CR is used as base classifier in the two level ensemble classifier approach. For experiments, two datasets, NSL-KDD and UNSW-NB15 are used. The proposed method reduced features from 41 to 37 and 49 to 19 over NSL KDD and UNSW-NB15 respectively.

In [20], the authors proposed an algorithm called PCANNA (Principal Component Analysis Neural Network Algorithm), which uses Principal Component Analysis for It reduced 41 features into 8 features. The experiments of this proposed method are done on NSLKDD datasets. In this proposed algorithm, PCA has been used to reduce the features, and the back propagation algorithms as learning tools. A Neural network has been used to identify unknown attacks. It reduced training time by 40\% and testing time by 78.5%. It also significantly reduced computer resource, memory and CPU time to detect attacks.

In [21], authors proposed a hybrid classification approach that used the combination of K mean clustering algorithm and RBF kernel function of Support Vector Machine (SVM). K mean cluster used to select the features. Features are selected based on different attacks types. KDDCUP 99 datasets are used to evaluate proposed hybrid approach. K mean cluster algorithm applied on training data then SVM with RBF kernel function applied as a classifier to detect different attacks class. Hybrid method K mean and SVM (KMSVM) increased the accuracy, detection rate as compared to separate K mean and SVM.

In [22], authors proposed a hybrid feature selection method that used filter based and wrapper-based methods. The Mutual information (MI) method was used as a filter-based approach and the highest MI score features were selected. Entropy measured the mutual information between random variables. Firefly algorithm with C4.5 and Bayesian Networks (BN) classifiers is used as a wrapper method to select the best features. The MIFA hybrid approach selected 10 important features. KDD CUP 99 dataset is used to assess the performance of the suggested technique.

## 3. Proposed Methodology

Proposed model is demonstrated in Figure 1. Model consists of following three phases:

- Preprocessing Phase
- Features selection phase
- Model & evaluation Phase

In preprocessing phase, we load the datasets for cleaning and normalizing. In the feature selection phase, the GA-based feature selection method is implemented on cleaned datasets. GA uses the NB algorithm as its fitness function and selects the most relevant feature sets. In the model and evaluation phase, two ensemble classifiers (stacking and bagging) train the datasets with a specific feature set that was generated in the previous phase. Trained model evaluates through different metrics. The architectural design of the suggested model is discussed in more detail in the next subsections.

### 3.1. Preprocessing phase

Preprocessing is the first and most important part of the ML field. The purpose of preprocessing is to clean and normalize the data. The data was cleaned by removing infinity values and filling the missing values. We fill the missing values by using the constant 0 in UNSW-NB15 dataset and removed infinity values by using the mean of column in CICDDoS2019. We used label encoding to convert categorical features into integers. Moreover, for normalization, we used the Min–Max scaling method. In the Min–Max method, data is scaled in the range of 0 and 1.

### 3.2. Feature selection phase

Feature selection is a machine learning technique that used to select best features from dataset and makes dataset lightweight, it can help to improve accuracy and detection rates and decrease computational time of IDS. In the feature selection phase, we implemented the wrapper-based GA to select relevant features. The genetic algorithm is explained in detail in the next subsection.

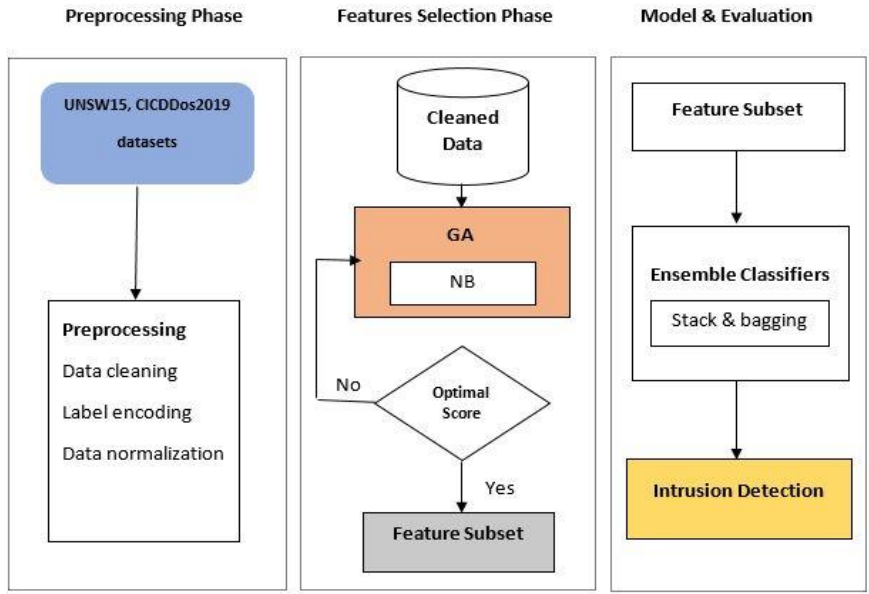

Figure 1. Phases of suggested model

### 3.2.1. Genetic algorithm

The Genetic Algorithm (GA) is an evolutionary algorithm that is used to solve various optimization problems. GA is a heuristic search-based algorithm that is inspired by the process of natural evolution. A genetic algorithm generates a new population by selecting individuals based on their fitness score. Then selected individuals recombined with other operators of the GA to generate the best solution. It makes different combinations of the inputs and finds the best combination based on each individual score [15]. GA calculates the fitness values of every individual and combines them with other operators to generate the best subset of features, and high fitness value individuals are considered as the best features subset.

The main steps of the GA algorithm are as follows [23]:

- Initialize the Population
- Calculate the fitness values
- Selection
- Crossover
- Mutation

**Initialize Population:** Population consist of set of individuals and every individual have set of gene (features). The gene of individuals in initial population normally randomly generated. In our problem every gene encoded in binary string, where 0 means that a feature with that index is not selected and 1 indicates that feature with that index is selected.

**Fitness Operator:** Fitness function is a basic element of genetic algorithms. After initialization, each individual is assigned a fitness value. Individuals with higher fitness scores have a higher chance of being selected for reproduction. We applied Naïve Bayes (NB) classifiers as a fitness function in our research. Naïve Bayes algorithm evaluated the performance of each individual and assigned the classification accuracy by using 5-fold cross validation. Individual values with higher accuracy are chosen for recombination.

**Selection:** This phase selects the individuals that are used for next generation. Individuals are chosen based on their fitness values. The selected individuals pass on their genes to next generation for create best solution than previous generation. There are three common methods used to selects individuals for recombination: Roulette wheel selection, Elitist selection and Tournament selection. We used 'Tournament selection' in our research. This selection process randomly selects the individuals from the population and applies a number of tournaments to select a highest fitness score individual for next generation.

**Crossover:** Select the half population of individuals in selection phase, crossover randomly selects individuals and combines their features to create the offspring for new generation. Crossover stops, when the new population reaches the same size as the old. In our research, we applied 'two point' approach that decides which features are taken from which individuals for creating offspring. It takes randomly two points and swapped the both individuals.

**Mutation:** Crossover operator can produce offspring that look very similar to parents and that cause the low diversity in new generation. Mutation add some randomness in offspring to avoid this problem. It changes some offspring by using different methods. We used the 'Flip Bit' approach, which randomly flips the bits in offspring to create diversity.

### 3.3. Model & Evaluation phase

For training and classification of the model, we implemented ensemble-based classifiers in our research. Ensemble methods are used to combine the predictions of various base estimators with given learning algorithms and improve the performance over single classifiers. The three main classifiers of ensemble approaches are bagging, boosting and stacking classifiers. We implemented stacking and bagging ensemble classifiers in our research.

**Stacking Classifier:** This ensemble method combines the estimator's prediction and takes these predictions as an input for final estimator. Stacking classifier consist of two level (level o and level 1). Level 0 computes the base classifiers prediction and level 1 takes these predictions as input and compute final prediction with meta classifier. We used three classifiers: Random Forest (RF), Decision Tree (DT), and naive bayes as base classifiers and RF as meta classifier. The meta classifier takes the predictions of these three classifiers as an input and computes the final prediction. We used random forest classifiers in our estimators and also in final estimators because they gave us high accuracy compared to other single classifiers.

**Bagging Classifier:** Bagging ensemble method make various independent models and aggregate their prediction. This method fits the base classifier on a random subset of the original dataset and aggregates each prediction to make a final prediction. We used the RF learning algorithm as the base classifier in our research.

**Hyperparameter Tuning**

Hyperparameter tuning is a machine learning approach to selecting a set of optimal classifier parameters. Grid search and Random search are the two most used approaches for hyperparameter tuning. We used the grid search approach to get the best RF and DT hyperparameter. We applied grid search on n_estimators, criterion, max_depth, max_features, min_samples_split and min_samples_leaf parameters and obtained on n_estimators is 100, criterion is 'gini', max_depth is 80, max_features is 2, min_samples_split is 8 and min_samples_leaf is 3. For DT we obtained n_estimators are 100, criterion is 'gini' and min_samples_split is 2.

**Performance Metrics**

Model evaluation helps to assess the performance of modal generalization. Most common evaluation metrics is accuracy. Accuracy is the number of correctly predicted data points from all data points. It used to verify how much model correctly work. Different types of metrics also used to evaluate the performance of model such as: precision, recall, false alarm rate and detection rate. We have used the following metrics to evaluated the performance of our model, Accuracy (ACC), Precision (PR), Recall (RC), F1-score (F1s), false alarm rate (FAR) and Detection rate (DR). Precision is number of true positives divided by the addition of total true positives and false positives. Accuracy of the modal is defined by diving correctly classified points by total points. Recall matrix is used to define the correctly measured true positive values that the model has detected. F1s, also known as F-score or F measure, is based on the balance between precision and the recall. FAR (False Alarm Rate) identifies the attack as normal traffic. We take DR as a correctly predicted attack.

Performance matrices are computed as follows:

$$ACC = \frac{(TP + TN)}{(TP + TN + FP + FN)} \quad (1)$$

$$RC = \frac{TP}{(TP + FN)} \quad (2)$$

$$PR = \frac{TP}{(TP + FP)} \tag{3}$$

$$F1s = 2\frac{RC \times PR}{RC + PR} \tag{4}$$

$$FAR = \frac{FN}{(FN + TP)} \tag{5}$$

$$DR = \frac{TP}{(TP + FN)} * 100 \tag{6}$$

Above mentioned matrices are defined as follows:

- True Positive (TP): True positives defines the correctly labeled intrusion attacks
- True Negative (TN): True negatives represent the correctly labeled normal or non-intrusive traces
- False Positive (FP): False positive represents the traces that are incorrectly labeled as attacks
- False Negative (FN): False negatives represent the traces that are incorrectly labeled as normal network traces (non-intrusive) when they are actually attacks.

## 4. Experiments & Discussion

Our research experiments conducted on two datasets UNSW-NB15 and CICDDoS2019. The UNSW-NB15 is one of the latest datasets that are publicly available. In IDS research, this dataset is widely used [16]. UNSW-NB15 dataset generated by the research group of Australian Centre of Cyber Security's (ACCS) for IDS model evaluation [24]. This dataset contains two files training and testing that divided into 175,341 and 82,332 respectively [24]. UNSW-NB15 datasets consist of 10 classes and 45 features where two features 'attack_cat' and 'label' used as target variable. Label used for binary classification and attack_cat used as multiclass classification. We dropped label feature and keep attack_cat for multiclass classification. UNSW-NB15 is an imbalance dataset, Worms,

Table 1. Records distribution of UNSW-NB15 & CICDDoS2019 datasets

| UNSW-NB15 | | CICDDoS2019 | |
|---|---|---|---|
| Class | No.of Records | Class | No.of Records |
| Normal | 2135 | UDPLag | 1873 |
| Fuzzers | 2112 | UDP | 2076 |
| Analysis | 2000 | Benign | 2071 |
| Dos | 2145 | Syn | 2027 |
| Exploits | 2146 | | |
| Reconnaissance | 2097 | | |
| Generic | 2081 | | |

Shellcodes, Analysis and Backdoors classes have extremely low records. In this research we used UNSW-NB15 training file (UNSW_NB15_training-set.csv). Some research studies [25,26] also use different dataset sizes randomly selected from the training dataset. We used dataset with 14797 records and with 7 classes. Table 1. depicts the records distribution of the dataset that we used in our research experiments.

CICDoS2019 is a most recent dataset that contains the most up to-date DDos attack types. In terms of feature selection in intrusion detection datasets, this dataset has not been used commonly by researchers. This dataset has 87 features and 12 attacks in training datasets and 7 attacks in testing dataset [12,27]. We used in our research UDPLag dataset file. The total number of records in this dataset is 7,24,770, and it includes four different types of DDos attacks classes namely, Benign. Syn, UDP and UDPLag. This is an imbalance dataset and majority of the records in this dataset belong to the Syn class. To avoid data imbalance, we reduced the number of records and select 8047 records. The distribution of both datasets records that we used in our research is shown in Table 1.

The experiments of the suggested methods were performed in Python using the Anaconda Navigator jupyter Notebook (version 6.1.4) on a 2.4 GHz Intel Core i5 CPU with 8 GB RAM.

The experimental results of stacking and bagging algorithms by using UNSW-NB15 are presented in Table 2 and Table 3 respectively. After feature selection process, we used ensemble classifier: stacking and bagging on selected features and we repeated this process 10 times with 10 cross validations. For evaluation of models, there are the following matrices we used to evaluate our results: Precision (PR), recall (RC), F1 scores, accuracy (ACC), FAR, and DR. Table 4 provides comparative results of our suggested models against existing methods.

The evaluation measures for each class of the UNSW-NB15 dataset by using the stacking classifier are listed in Table. 2. The results show that ensemble classifiers detected all classes. It shows that the normal and generic class have high DR and low FAR, where the DR and FAR for the normal class are 98.93 and 0.01 respectively, and for the generic class are 98.94 and 0.01 respectively. Moreover, results also show that the DoS class has low DR and high FAR that is 69.70 and 0.30 respectively. Overall, this ensemble method performed in other classes is also good.

The evaluation of each class using the bagging classifier is represented in Table 3. where the generic and normal classes have high detection rate and low FAR. Generic class DR is 99.82 and FAR is 0.010, and for normal class the DR is 96.73 and FAR is 0.029. Overall, the bagging ensemble classifier also performed better in other classes.

Table 2. Evaluation measures for each class of UNSW-NB15 by using stacking algorithm

| Class Name | PR | RC | F1s | FAR | DR |
|---|---|---|---|---|---|
| Analysis | 0.893 | 0.8122 | 0.853 | 0.17 | 81.48 |
| Dos | 0.708 | 0.686 | 0.699 | 0.30 | 69.7 |
| Exploits | 0.663 | 0.835 | 0.739 | 0.16 | 83.15 |
| Fuzzers | 0.871 | 0.89 | 0.88 | 0.10 | 89.55 |
| Generic | 0.992 | 0.99 | 0.99 | 0.01 | 98.93 |
| Normal | 0.993 | 0.984 | 0.988 | 0.01 | 98.94 |
| Reconnaissance | 0.923 | 0.785 | 0.85 | 0.21 | 78.52 |

Table 3: Evaluation measures for each class of UNSW-NB15 using bagging algorithm

| Class Name | PR | RC | F1s | FAR | DR |
|---|---|---|---|---|---|
| Analysis | 0.892 | 0.831 | 0.862 | 0.168 | 83.12 |
| Dos | 0.591 | 0.728 | 0.652 | 0.277 | 72.81 |
| Exploits | 0.713 | 0.683 | 0.695 | 0.330 | 68.34 |
| Fuzzers | 0.851 | 0.835 | 0.845 | 0.138 | 83.52 |
| Generic | 0.99 | 0.99 | 0.99 | 0.010 | 99.82 |
| Normal | 0.991 | 0.967 | 0.981 | 0.029 | 96.73 |
| Reconnaissance | 0.832 | 0.768 | 0.8 | 0.213 | 76.85 |

Table 4. Comparison with other methods on UNSW-NB15

| Ref | FS Method | Classifiers | PR | RC | F1s | ACC | FAR | DR% |
|---|---|---|---|---|---|---|---|---|
| [8] | XGBoost(gain) | XGBoost | 0.78 | 0.75 | 0.76 | 0.7576 | 0.245 | 75.45 |
| [28] | GA-LR | DT | 0.62 | 0.6 | 0.61 | 0.731 | 0. 267 | 0.731 |
| [16] | GA | ET | 0.74 | 0.72 | 0.73 | 0.730 | - | - |
| | | RF | 0.66 | 0.65 | 0.65 | 0.64 | | |
| | | DT | 0.754 | 0.727 | 0.732 | 0.725 | | |
| | | NB | 0.51 | 0.472 | 0.48 | 0.47 | | |
| | | XGB | 0.739 | 0.720 | 0.724 | 0.72 | | |
| [11] | XGBoost | KNN | 0.754 | 0.744 | 0.744 | 0.742 | - | - |
| | | DT | 0.718 | 0.712 | 0.712 | 0.71 | | |
| | | SVM | 0.674 | 0.624 | 0.622 | 0.62 | | |
| | | LR | 0.596 | 0.572 | 0.568 | 0.57 | | |
| | | ANN | 0.674 | 0.66 | 0.648 | 0.65 | | |
| **Suggested Model** | GA | Stacking (RF,NB,DT) | 0.864 | 0.856 | 0.856 | 0.856 | 0.136 | 85.75 |
| **Suggested Model** | GA | Bagging (RF) | 0.837 | 0.828 | 0.832 | 0.832 | 0.166 | 82.88 |

The comparison of the suggested model on UNSW-NB15 with other studies is described in Table 4. In terms of accuracy, PR, RC, FAR and DR the suggested technique achieved encouraging results. Our suggested stacking method obtained a 10.44% and bagging obtained an 8.04% higher accuracy compared to the method presented in [8]. Also, our method performs better in terms of FAR and DR. In [28], authors proposed GA-LR method for feature selection with DT classifier, in comparison with this method, our suggested method obtained 12.5% & 25.51% higher accuracy and DR respectively.

In [16], the authors proposed a model that used GA for feature selection and applied different classifiers, where extra tree (ET) obtained 0.73 highest accuracy compared to other classifiers. Compared to method provided in [16], our suggested stacking and bagging methods obtained 12.56% and 10.16% higher accuracy, respectively.

In terms of PR, RC, F1s, and ACC, our suggested work performed higher than the method presented in [11], where maximum accuracy of 0.74 is obtained by the KNN classifier. The stacking model is higher with 11.4% accuracy, 11% PR, 11.2% RC and 11.2% F1s, and the bagging method is higher with 9, 8.3, 8.4 and 8.8% accuracy, PR, RC and F1s respectively.

The comparison of all classifiers of our suggested model is demonstrated in Figure 2. and it shows that the ensemble classifiers achieved better accuracy than single classifiers. In single classifiers, RF performed better than NB and DT, and in ensemble methods, stacking performed better than bagging.

The experiments performed on CICDDoS2019 using stacking and bagging algorithms are described in Tables 5 and Table 6 respectively, while Table 7 presents results of the suggested model as compared to existing classifiers.

Table 5. Evaluation measures for each class of CICDDoS2019 by using stacking algorithm

| Class | PR | RC | F1s | FAR | DR |
|---|---|---|---|---|---|
| BENIGN | 0.991 | 1 | 0.999 | 0.001 | 99.86 |
| SYN | 1 | 1 | 1 | 0.0004 | 99.99 |
| UDP | 1 | 1 | 1 | 0.0008 | 99.91 |
| UDPLag | 1 | 0.991 | 0.994 | 0.006 | 99.37 |

Table 6: Evaluation of each class of CICDDoS2019 using bagging algorithm

| Class | PR | RC | F1s | FAR | DR |
|---|---|---|---|---|---|
| BENIGN | 0.98 | 1 | 0.99 | 0.00264 | 99.76 |
| SYN | 1 | 1 | 1 | 0.00042 | 99.99 |
| UDP | 1 | 1 | 1 | 0.00235 | 99.91 |
| UDPLag | 0.999 | 0.974 | 0.982 | 0.02818 | 97.42 |

Table 7: Comparison of suggested models with other models using CICDDoS2019 dataset

| Ref | FS Method | Classifiers | PR | RC | F1s | ACC | FAR | DR |
|---|---|---|---|---|---|---|---|---|
| [12] | Info gain | ID3 | 0.97 | 0.98 | 0.97 | - | - | - |
| | | RF | 0.99 | 0.99 | 0.99 | - | - | - |
| | | NB | 0.98 | 0.98 | 0.98 | - | - | - |
| | | LR | 0.98 | 0.98 | 0.98 | - | - | - |
| **Suggested Models** | GA | Stacking | 0.997 | 0.997 | 0.999 | 0.997 | 0.00177 | 99.783 |
| | GA | Bagging | 0.99 | 0.991 | 0.99 | 0.993 | 0.00829 | 99.7402 |

The evaluation measures for each class of the CICDDos2019 dataset for both stacking and bagging ensemble classifiers are listed in Table 5 & 6 respectively. The results show that the ensemble classifiers detected all classes. It shows that all classes have high DR and low FAR, although in both ensemble methods, the Syn class has high DR and low FAR in all of the classes that are 99.99 and 0.0004 respectively. Overall, the ensemble methods performed in other classes is also very good. Table. 7 shows the overall evaluation performance of our proposed methods on CICDoS2019 dataset.

The comparison with the study in [12] shows that our suggested approach performed better in terms of PR, RC, and F1. Results show that the stacking algorithm performed better than the bagging algorithm. The stacking algorithm achieved the highest score in all the evaluation metrics. Figure 2. illustrates the overall classifier performance, where it shows that ensemble methods performed very well compared to single classifiers. In single classifiers, RF performed better compared to NB and DT, and in ensemble methods, stacking classifier achieved high accuracy.

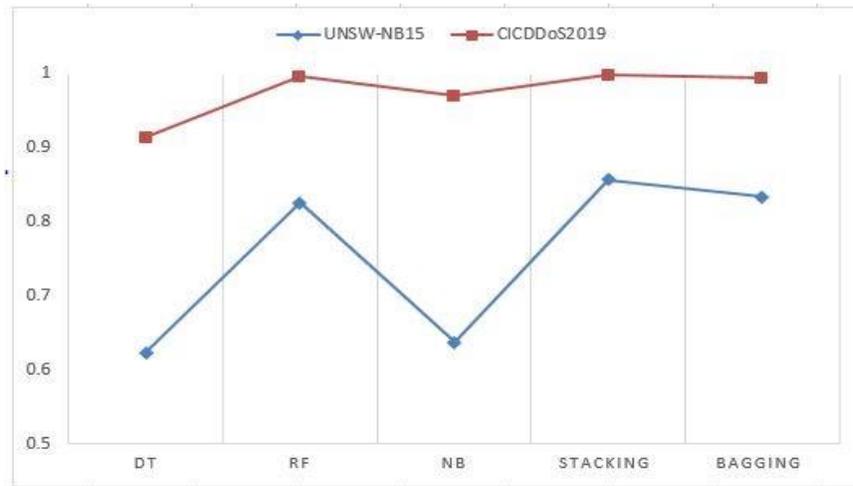

Figure 2. Evaluation of classifiers using both datasets

## 5. Conclusion

Our research is based on intrusion detection system to detect different types of attacks. To achieve high accuracy and low FAR there should be light weight dataset. Features selection process reduce the unnecessary features and improve the accuracy. We suggested GA based features selection method that achieved high accuracy, DR and low FAR. In our research, GA used as a feature selection method and ensemble classifiers used for classification of different attack types. We used two ensemble classifiers: stacking and bagging classifier. UNSW-NB15 and CICDDoS2019 are the two multiclass datasets used in this research. Using UNSW-NB15, we achieved 0.856 accuracy and 0.136 FAR by using a stacking classifier. In UNSW-NB15 dataset the DR of normal class is 98.93, Generic 98.94, Analysis 81.48. Dos 69.70, Exploits 83.15, Fuzzers 89.55 and Reconnaissance 78.52 achieved. Using CICDDoS2019 dataset, we obtained 0.997 accuracy and 0.00177 FAR. In the CICDDoS2019 dataset, four classes were used for multiclass classification, and the detection rate for the UDP class is 99.91, UDPLag is 99.37, syn 99.99, and Benign is 99.83. Stacking and bagging ensemble classifiers improved the accuracy compared to other existing methods.

# References

1. *A survey on feature selection for intrusion detection.* **Zuech R, Khoshgoftaar TM.** 2015. In Proceedings of the 21st ISSAT International Conference on Reliability and Quality in Design. pp. 150-155.

2. *A study of feature selection methods in intrusion detection system: a survey.* **P, Ahmed.** 2012 Sep, International Journal of Computer Science Engineering and Information Technology Research [IJCSEITR], ISSN No: 2249–7943., Vol. 2(3), pp. 1-25.

3. *Feature selection techniques for intrusion detection using non-bio-inspired and bio-inspired optimization algorithms.* **Balasaraswathi VR, Sugumaran M, Hamid Y.** 2017 Dec 1, Journal of Communications and Information Networks, Vols. 107-19, p. 2(4).

4. *Intrusion detection with comparative analysis of supervised learning techniques and fisher score feature selection algorithm.* **Aksu D, Üstebay S, Aydin MA, Atmaca T.** Springer, 2018 Sep 20, In International symposium on computer and information sciences, pp. pp. 141-149.

5. *Unsupervised feature selection using feature similarity.* **Mitra P, Murthy CA, Pal S.** 2002 Aug 7, IEEE transactions on pattern analysis and machine intelligence, Vol. 24(3), pp. 301-12.

6. *Intrusion detection and big heterogeneous data: a survey.* **Zuech R, Khoshgoftaar TM, Wald R.** 2015 Dec, Journal of Big Data, Vol. 2(1), pp. 1-41.

7. *Toward integrating feature selection algorithms for classification and clustering.* **Liu H, Yu L.** s.l. : IEEE, 2005 Mar 7, IEEE Transactions on knowledge and data engineering, Vol. 17(4), pp. 491-502.

8. *Development of an efficient network intrusion detection model using extreme gradient boosting (XGBoost) on the UNSW-NB15 dataset.* **Husain A, Salem A, Jim C, Dimitoglou G.** 2019 Dec 10, In2019 IEEE International Symposium on Signal Processing and Information Technology (ISSPIT), pp. 1-7.

9. *Feature ranking and selection for intrusion detection using artificial neural networks and statistical methods.* **Tamilarasan A, Mukkamala S, Sung AH, Yendrapalli K.** s.l. : IEEE, 2006 Jul 16 . InThe 2006 IEEE International Joint Conference on Neural Network Proceedings. pp. 4754-4761.

10. *A two-stage classifier approach using reptree algorithm for network intrusion detection.* **Belouch M, El Hadaj S, Idhammad M.** 2017 Jul, International Journal of Advanced Computer Science and Applications, Vol. 8(6), pp. 389-94.

11. *Performance analysis of intrusion detection systems using a feature selection method on the UNSW-NB15 datase.* **}Kasongo SM, Sun Y.** 2020 Dec, Journal of Big Data, Vol. 7(1), pp. 1-20.

12. *Developing realistic distributed denial of service (DDoS) attack dataset and taxonomy.* **Sharafaldin I, Lashkari AH, Hakak S, Ghorbani AA.** s.l. : IEEE, 2019 Oct 1 . In2019 International Carnahan Conference on Security Technology (ICCST). pp. 1-8.

13. *Selection and performance analysis of CICIDS2017 features importance.* **Reis B, Maia E, Praça I.** s.l. : Springer, 2019 Nov 5 , InInternational Symposium on Foundations and Practice of Security, pp. 56-71.

14. *A NSGA2-LR wrapper approach for feature selection in network intrusion detection.* **Khammassi C, Krichen S.** 2020 May 8, Vol. 172:107183.

15. *A new feature selection IDS based on genetic algorithm and SVM.* **Gharaee H, Hosseinvand H.** 2016 Sep 27 , In2016 8th International Symposium on Telecommunications (IST), pp. 139-144.

16. *An Advanced Intrusion Detection System for IIoT Based on GA and Tree Based Algorithms.* **SM, Kasongo.** 2021 Aug , IEEE Access, Vol. 11;9, pp. 113199-212.

17. *A feature selection algorithm for intrusion detection system based on pigeon inspired optimizer.* **Alazzam H, Sharieh A, Sabri KE.** 2020 Jun 15, Expert systems with applications, p. 148:113249.

18. *Building an efficient intrusion detection system based on feature selection and ensemble classifie.* **Zhou Y, Cheng G, Jiang S, Dai M.** 2020 Jun 19, Computer networks, p. 174:107247.

19. *TSE-IDS: A two-stage classifier ensemble for intelligent anomaly-based intrusion detection system.* **Tama BA, Comuzzi M, Rhee KH.** 2019 Jul 11, IEEE Access, Vol. 7, pp. 94497-507.

20. *Feature reduction using principal component analysis for effective anomaly–based intrusion detection on NSL-KDD.* **Lakhina S, Joseph S, Verma B.** 2010 july, International Journal of Engineering Science and Technology .

21. *Feature selection based hybrid anomaly intrusion detection system using K means and RBF kernel function.* **}Ravale U, Marathe N, Padiya P.** 2015 Jan , Procedia Computer Science, Vol. 1;45, pp. 428-35.

22. *Firefly algorithm based feature selection for network intrusion detection.* **Selvakumar B, Muneeswaran K.** 2019 Mar 1, Computers & Security, Vol. 81, pp. 148-55.

23. *An improved intrusion detection algorithm based on GA and SVM.* **Tao P, Sun Z, Sun Z.** 2018 Mar 5, IEEE Access, Vol. 6, pp. 13624-31.

24. **UNB Canadian Institute for Cybersecurity. [Online] https://www.unb.ca/cic/datasets/nsl.html.**

25. *A GA-LR wrapper approach for feature selection in network intrusion detection.* **Khammassi C, Krichen S.** 2017 Sep 1, computers & security. , Vol. 70, pp. 255-77.

26. *A novel feature-selection approach based on the cuttlefish optimization algorithm for intrusion detection systems.* **Eesa AS, Orman Z, Brifcani AM**. 2015 Apr 1, Expert systems with applications, Vol. 42(5), pp. 2670-9.

27. *Tensor based framework for Distributed Denial of Service attack detection.* **Maranhão JP, da Costa JP, Javidi E, de Andrade CA, de Sousa Jr RT.** 2021 Jan 15, Journal of Network and Computer Applications, Vol. 174, p. 102894.

28. *A GA-LR wrapper approach for feature selection in network intrusion detection.* **Khammassi C, Krichen S.** s.l. : Elsevier , 2017 Sep 1, computers & security, Vol. 70, pp. 255-77.

29. *An improved intrusion detection algorithm based on GA and SVM.* **Tao P, Sun Z, Sun Z.** 2018 Mar 5, Ieee Access, Vol. 6, pp. 13624-31.